\documentclass[12pt]{article}
\usepackage[cp1251]{inputenc}
\usepackage{amsmath}
\usepackage{amsfonts}
\usepackage{amssymb}
\def\pa{\partial}                       
\def\beq{\begin{eqnarray}}    
\def\eeq{\end{eqnarray}}      

\def\sDet{\,\mbox{sDet}\,}              

\begin{document}
\date{}
\begin{center}
{\Large\textbf{Anticanonical transformations and  Grand Jacobian}}

\vspace{18mm}

{\large I.A. Batalin$^{(a)}\footnote{E-mail:
batalin@lpi.ru}$\;,
P.M. Lavrov$^{(b)} \footnote{E-mail:
lavrov@tspu.edu.ru}$,\;
I.V. Tyutin$^{(a)}\footnote{E-mail:
tyutin@lpi.ru}$\;
}

\vspace{8mm}

\noindent ${{}^{(a)}}$
{\em P.N. Lebedev Physical Institute,\\
Leninsky Prospect \ 53, 119 991 Moscow, Russia}

\noindent  ${{}^{(b)}}
${\em
Tomsk State Pedagogical University,\\
Kievskaya St.\ 60, 634061 Tomsk, Russia}

\vspace{20mm}

\begin{abstract}
\noindent An independent (purely algebraic) proof of
factorization property of the Grand Jacobian corresponding to
anticanonical transformations in the BV-formalism is given.
\end{abstract}

\end{center}

\vfill

\noindent {\sl Keywords:} BV-formalism, BRST symmetry,
anticanonical transformations
\\

\noindent PACS numbers: 11.10.Ef, 11.15.Bt
\newpage

\section{Introduction}

The BV formalism (or the field-antifield formalism) \cite{BV,BV1} is a powerful covariant
quantization method which can be applied to arbitrary gauge
invariant systems. This method is based on fundamental concept
of global supersymmetry known as the BRST symmetry
\cite{brs1,t}.
One of the most important objects of the field-antifield
formalism is an odd symplectic structure called antibracket and
known to mathematicians  as Buttin bracket \cite{Butt}. In terms of
antibracket the classical master equation and the Ward identity for generating
functional of vertex functions (effective action) are formulated. It
is an important property that the antibracket is preserved under the
anticanonical transformations which are dual to canonical
transformations for a Poisson bracket. An important role and rich
geometric possibilities of general anticanonical transformations in
the field-antifield formalism have been realized in the procedure of
gauge fixing \cite{VLT}. Original procedure
of gauge fixing \cite{BV,BV1} corresponds in fact to a special type
of anticanonical transformation in an action being a proper solution
to the quantum master equation.

Anticanonical transformations play
a crucial role in describing the structure of renormalization and
gauge dependence of the effective action in general gauge theories
\cite{VLT}. Another important application  is
the study of  arbitrariness in solutions to the classical master equation
\cite{LT} and to the quantum master equation \cite{BL,BLT} when the Grand Jacobian
of anticanonical transformations presents as an essential part of full quantum action.
The Grand Jacobian possesses an interesting property known as the factorization property
allowing to present it through the superdeterminant of supermatrix in the sector
of anti canonically transformed fields only. Other possibility is related to use in this
presentation of the Grand Jacobian the superdeterminant of supermatrix in the sector
of anti canonically transformed antifields. These properties of the Grand Jacobian
were known at least since the article \cite{BV2}, although the proof was omitted therein.
Later on we filled this gap by proving the factorization property of the Grand Jacobian
with the help of a solution to the Lie equation for one-parameter
family of antisymplectic variables subjected to anticanonical transformations \cite{BLT}.

At the present paper we are going to give  a simple proof of the
factorization property of the Grand Jacobian corresponding to  anticanonical
transformations within the field-antifield formalism \cite{BV,BV1} based on
using algebraical properties of anticanonical transformations only.

We use the DeWitt's condensed notations \cite{DeWitt}.
We employ the notation $\varepsilon(A)$ for the Grassmann parity of
any quantity $A$.  The functional derivatives with respect to fields and antifields
are considered as  left ones.   The right
functional derivatives  are marked by special symbol $"\leftarrow"$.

\section{Anticanonical transformations}

We will proceed with the use of antisymplectic  Darboux coordinates
$z^{A}$ in the form of an explicit splitting  into  fields $\phi^{i}$
and antifields $\phi^*_{i}$,
 \beq
 \label{E1}
 z^{A} = \{\phi^{i},\phi^*_{i}\},\quad \varepsilon(z^A)=\varepsilon_A,\quad
\varepsilon(\phi^*_i)=\varepsilon(\phi^i)+1.
\eeq
For any functionals $G=G(\phi,\phi^*), H=H(\phi,\phi^*)$ the antibracket
is defined by the rule
\beq
 \label{E2}
(G,H)=G\Big(\frac{\overleftarrow{\pa}}{\pa\phi^i}\frac{\pa}{\pa \phi^*_i}-
\frac{\overleftarrow{\pa}}{\pa\phi^*_i}\frac{\pa}{\pa \phi^i}\Big)H,\quad
\varepsilon\big((G,H)\big)=\varepsilon(G)+\varepsilon(H)+1,
\eeq
so that
\beq
\label{E3}
 (z^A,z^B)=E^{AB},\quad \varepsilon\big(E^{AB}\big)=\varepsilon_A+\varepsilon_B+1,
\eeq
where $E^{AB}$ are elements of  a constant invertible antisymplectic metric $E$ with the
following block structure
\beq
\label{E4}
E=\left(\begin{array}{cc}
0 &I\\
-I & 0\\
\end{array}\right),\quad
E^{AB}=\left(\begin{array}{cc}
0 & \delta^i_{\;\!j}\\
-\delta^i_{\;\!j} & 0\\
\end{array}\right)
\eeq
and antisymmetry property
\beq
\label{E5}
 E^{AB}=-(-1)^{(\varepsilon_A+1)(\varepsilon_B+1)}E^{BA}.
\eeq
In terms of $z^A$ the antibracket rewrites as
\beq
\label{E6}
(G,H)=G\Big(\frac{\overleftarrow{\pa}}{\pa z^A}
E^{AB}\frac{\pa}{\pa z^B}\Big)H
\eeq

Let $F = F(\phi,\Phi^*)$ , $\varepsilon(F)=1$ be a generator of the
anticanonical transformation,
\beq
\label{E7}
\Phi^i=\frac{\pa}{\pa
\Phi^*_i} F(\phi,\Phi^*),\quad \phi^*_i=
\frac{\pa}{\pa \phi^i}F(\phi,\Phi^*).
\eeq
Any anticanonical transformation preserves the antibracket,
\beq
\label{E8}
(Z^A,Z^B)=E^{AB},
\eeq
where $Z^A=(\Phi^i,\Phi^*_i)$,
$\varepsilon(Z^A)=\varepsilon(z^A)=\varepsilon_A$
are considered as the functions of $z^A$, $Z^A=Z^A(z)$, found
from equations (\ref{E2}).
Condition of solvability for the anticanonical transformation leads to the
following relations
\beq
\label{E9}
&& \Big(Z^A(z)\frac{\overleftarrow{\pa}}{\pa z^C}\Big)
 \Big(z^C(Z) \frac{\overleftarrow{\pa}}{\pa Z^B}\Big)=\delta^{A}_{\;B}\;,
\quad
\Big(z^A(Z)\frac{\overleftarrow{\pa}}{\pa Z^C}\Big)
\Big(Z^C(z)\frac{\overleftarrow{\pa}}{\pa z^B}\Big)=\delta^{A}_{\;B}.
\eeq

Let $H^{A}_{\;B}$ be elements of the supermatrix $H$ of anticanonical transformation
(\ref{E7})
\beq
\label{E10}
H^{A}_{\;B}=\Big(Z^A(z)\frac{\overleftarrow{\pa}}{\pa z^B}\Big),\quad
\varepsilon(H^{A}_{\;B})=\varepsilon_A+\varepsilon_B,
\eeq
and
\beq
\label{E11}
H=
\left(\begin{array}{cc}
{\cal A} & {\cal B}\\
{\cal C} & {\cal D} \\
\end{array}\right), \qquad
H^{A}_{\;B}=
\left(\begin{array}{cc}
{\cal A}^{i}_{\;j}&{\cal B}^{ij}\\
{\cal C}_{ij}&{\cal D}_{i}^{\;j}\\
\end{array}\right).
\eeq
Here the following notations
\beq \label{E12}
&&{\cal A}^{i}_{\;j}=\Phi^{i}(\phi,\phi^*)\frac{\overleftarrow{\pa}}{\pa\phi^j},
\qquad\;
{\cal B}^{ij}=\Phi^{i}(\phi,\phi^*)\frac{\overleftarrow{\pa}}{\pa \phi^*_j},\\
\label{E13}
&&{\cal C}_{ij}=\Phi^*_{i}(\phi,\phi^*)\frac{\overleftarrow{\pa}}{\pa\phi^j},
\qquad {\cal D}_{i}^{\;j}= \Phi^*_{i}(\phi,\phi^*)
\frac{\overleftarrow{\pa}}{\pa\phi^*_j}\;.
\eeq
are used. The quantities in (\ref{E12}) and (\ref{E13}) have the following assignment
of Grassmann parities
\beq
\label{E14}
&&\varepsilon({\cal A}^{i}_{\;j})=\varepsilon_i+\varepsilon_j,\qquad
\varepsilon({\cal B}^{ij})=\varepsilon_i+\varepsilon_j+1,\\
\label{E15}
&&\varepsilon({\cal C}_{ij})=\varepsilon_i+\varepsilon_j+1,\quad
\varepsilon({\cal D}_{i}^{\;j})=\varepsilon_i+\varepsilon_j\;.
\eeq
In terms of the anticanonical transformation we have
\beq
\nonumber
{\cal A}^{i}_{\;j}&=&\frac{\pa}{\pa\Phi^*_i}\frac{\pa}{\pa\phi^j}F+
\Big(\frac{\pa}{\pa\Phi^*_i}\frac{\pa}{\pa\Phi^*_k}F\Big){\cal C}_{kj}= \\
\label{E16}
&=&\frac{\pa}{\pa\Phi^*_i}\frac{\pa}{\pa\phi^j}F-
\Big(\frac{\pa}{\pa\Phi^*_i}\frac{\pa}{\pa\Phi^*_k}F \Big){\cal D}_{k}^{\;\;l}\Big(
\frac{\pa}{\pa\phi^l}\frac{\pa}{\pa\phi^j}F\Big), \\
\label{E17}
{\cal B}^{ij}&=&\left(\frac{\pa}{\pa\Phi^*_i}\frac{\pa}{\pa\Phi^*_k}F\right)
\left(\Phi^*_{k}(\phi,\phi^*)
\frac{\overleftarrow{\pa}}{\pa\phi^*_j}\right)=
\Big(\frac{\pa}{\pa\Phi^*_i}\frac{\pa}{\pa\Phi^*_k}F\Big){\cal D}_{k}^{\;\;j},
\eeq
or using the notations
\beq
\label{E18}
{\cal M}_{i}^{\;j}=\frac{\pa}{\pa\Phi^*_i}\frac{\pa}{\pa\phi^j}F,\quad
{\cal K}^{ij}=\frac{\pa}{\pa\Phi^*_i}\frac{\pa}{\pa\Phi^*_j}F,
\eeq
in short as
\beq
\label{E19}
{\cal A}={\cal M}+{\cal K}{\cal C},\quad {\cal B}={\cal K} {\cal D}.
\eeq

Let us introduce the supermatrix $R$ with elements $R^A_{\;B}$ as
\beq
\label{E20}
R^{A}_{\;B}=z(Z)^A\frac{\overleftarrow{\pa}}{\pa Z_B},\quad
\varepsilon(R^{A}_{\;B})=\varepsilon_A+\varepsilon_B
\eeq
and
\beq
\label{E21}
R^{A}_{\;B}=
\left(\begin{array}{cc}
{\cal V}^{i}_{\;j}&{\cal W}^{ij}\\
{\cal Z}_{ij}&{\cal U}_{i}^{\;j}\\
\end{array}\right),\quad
R=
\left(\begin{array}{cc}
{\cal V}&{\cal W}\\
{\cal Z}&{\cal U}\\
\end{array}\right)\;,
\eeq
where the quantities
\beq
\label{E22}
&&{\cal V}^{i}_{\;j}=\phi^{i}(\Phi,\Phi^*)\frac{\overleftarrow{\pa}}{\pa\Phi^j},
\quad\quad
{\cal W}^{ij}=\phi^{i}(\Phi,\Phi^*)\frac{\overleftarrow{\pa}}{\pa\Phi^*_j}\;,\\
\label{E23}
&&{\cal Z}_{ij}=\phi^*_{i}(\Phi,\Phi^*)\frac{\overleftarrow{\pa}}{\pa\Phi^j}\;,
\qquad\;\;\;
{\cal U}_{i}^{\;j}=\phi^*_{i}(\Phi,\Phi^*)\frac{\overleftarrow{\pa}}{\pa\Phi^*_j}\;.
\eeq
obey the following assignment  of Grassmann parities   
\beq
\label{E24}
&&\varepsilon({\cal V}^{i}_{\;j})=\varepsilon_i+\varepsilon_j,\;\quad
\varepsilon({\cal W}^{ij})=\varepsilon_i+\varepsilon_j+1,\\
\label{E25}
&&
\varepsilon({\cal Z}_{ij})=\varepsilon_i+\varepsilon_j+1,\quad
\varepsilon({\cal U}_{i}^{\;j})=\varepsilon_i+\varepsilon_j\;.
\eeq
In terms of anticanonical transformation we have
\beq
\nonumber
{\cal U}_{i}^{\;j}&=&(-1)^{(\varepsilon_j+1)\varepsilon_i}\Big(\frac{\pa}{\pa\Phi^*_j}
\frac{\pa}{\pa\phi^i}F\Big)+\Big(\frac{\pa}{\pa\phi^i}
\frac{\pa}{\pa\phi^k}F\Big)
\Big(\phi^k(\Phi,\Phi^*)\frac{\overleftarrow{\pa}}{\pa \Phi^*_j}\Big)=\\
\label{E26}
&=&\widetilde{{\cal M}}_i^{\;j}+{\cal N}_{ik}{\cal W}^{kj},\\
\label{E27}
{\cal Z}_{ij}&=&\Big(\frac{\pa}{\pa\phi^i}\frac{\pa}{\pa\phi^k}F\Big)
\Big(\phi^{k}(\Phi,\Phi^*)\frac{\overleftarrow{\pa}}{\pa\Phi^j}\Big)=
{\cal N}_{ik}{\cal V}^k_{\;j}\;,
\eeq
where $\widetilde{{\cal M}}$ is the supermatrix transposed with ${\cal M}$ (\ref{E18}),
\beq
\label{E28}
&&\widetilde{{\cal M}}_{i}^{\;j}=(-1)^{\varepsilon_i(\varepsilon_j+1)}{\cal M}_{j}^{\;i}=
\Big(\frac{\pa}{\pa\phi^i}\frac{\pa}{\pa\Phi^*_j}F\Big),\quad
\big(\widetilde{{\cal M}}^{-1}\big)_{i}^{\;j}=
(-1)^{\varepsilon_i(\varepsilon_j+1)}(M^{-1})^{j}_{\;i}\;,
\eeq
and
\beq
\label{E29}
{\cal N}_{ij}=\Big(\frac{\pa}{\pa\phi^i}\frac{\pa}{\pa\phi^j}F\Big)\;.
\eeq
From the first in (\ref{E9}) and (\ref{E11}), (\ref{E21}) we have
\beq
\label{E30}
&&{\cal A}{\cal V}+{\cal B}{\cal Z}=I\;,\qquad
{\cal A}{\cal W}+{\cal B}{\cal U}=0\;,\\
\label{E31}
&&{\cal C}{\cal V}+{\cal D}{\cal Z}=0\;,\qquad
{\cal C}{\cal W}+{\cal D}{\cal U}=I\;.
\eeq
From the first in (\ref{E31}) and (\ref{E27}) we derive the relation
\beq
\label{E32}
{\cal C}=-{\cal D} {\cal N},
\eeq
and therefore the presentation for ${\cal A}$ holds
\beq
\label{E33}
{\cal A}={\cal M}-{\cal K}{\cal D} {\cal N}.
\eeq
From the seconds in (\ref{E30}) and (\ref{E31}) it follows
\beq
\label{E34}
{\cal U}^{-1}={\cal D}-{\cal C}{\cal A}^{-1}{\cal B}\;.
\eeq
From the second in (\ref{E9}) and (\ref{E11}), (\ref{E21}) we have
\beq
\label{E35}
&&{\cal V}{\cal A}+{\cal W}{\cal C}=I\;,\qquad
{\cal V}{\cal B}+{\cal W}{\cal D}=0\;,\\
\label{E36}
&&{\cal Z}{\cal A}+{\cal U}\;\!{\cal C}=0\;,\qquad
{\cal Z}{\cal B}+{\cal U}\;\!{\cal D}=I\;.
\eeq
From the seconds in (\ref{E19})and (\ref{E35}) it follows
\beq
\label{E37}
{\cal W}=-{\cal V}{\cal K}.
\eeq
Using this relation from the first in (\ref{E35}) and  the second in (\ref{E36})
we derive the relations
\beq
\label{E38}
{\cal V}={\cal M}^{-1},\quad  {\cal D}=\widetilde{{\cal M}}^{-1},
\eeq
and the presentations for ${\cal A}$ and ${\cal U}$ in the form
\beq
\label{E39}
{\cal A}={\cal M}-{\cal K}\widetilde{{\cal M}}^{-1}{\cal N},\quad
{\cal U}=\widetilde{{\cal M}}-{\cal N}{\cal M}^{-1}{\cal K}.
\eeq
Consider now
\beq
\nonumber
&&\widetilde{(K\widetilde{M}^{-1}N)_i^{\;j}}=
(-1)^{(\varepsilon_j+1)\varepsilon_i}
(K\widetilde{M}^{-1}N)_{ji}=\\
\nonumber
&&=(-1)^{(\varepsilon_j+1)\varepsilon_i}
\frac{\pa^2 F}{\pa\Phi^*_j\pa\Phi^*_k}(\widetilde{M}^{-1})_{k}^{\;\;l}
\frac{\pa^2 F}{\pa\phi^l\pa\phi^i}=\\
\nonumber
&&=\frac{\pa^2 F}{\pa\phi^i\pa\phi^l}
(\widetilde{M}^{-1})_{k}^{\;\;l}(-1)^{(\varepsilon_l+1)\varepsilon_k}
\frac{\pa^2 F}{\pa\Phi^*_k\pa\Phi^*_j}=\\
\label{E40}
&&=\frac{\pa^2 F}{\pa\phi^i\pa\phi^l}
(M^{-1})^{l}_{\;\;k}
\frac{\pa^2 F}{\pa\Phi^*_k\pa\Phi^*_j}=(NM^{-1}K)_{i}^{\;j}\;.
\eeq
Therefore we obtain the relations
\beq
\label{E41}
{\cal A}=\widetilde{{\cal U}},\qquad{\cal U}=\widetilde{{\cal A}},
\eeq
playing a crucial role in proving the factorization property of the Grand Jacobian.

\section{Grand Jacobian}

Let $J$  be the Grand Jacobian  of anticanonical transformation
(\ref{E3}) which is expressed through the superdeterminant of
supermatrix $H$ (\ref{E10}),
\beq
\label{E42}
J=\sDet H.
\eeq
It is
known \cite{Berezin} that the superdeterminant of $H$ can be written
through the superdeterminants of its super blokes ${\cal A},{\cal
B},{\cal C},{\cal D}$ in (\ref{E11}) as
\beq
\label{E43}
\sDet H=(\sDet {\cal A})\sDet {\cal X}^{-1},\quad {\cal X}={\cal D}-{\cal
C}{\cal A}^{-1}{\cal B}.
\eeq
According to the relation (\ref{E34})
${\cal X}^{-1}={\cal U}$ so that taking into account (\ref{E41}) we
derive the equality
\beq
\label{E44}
J=(\sDet {\cal A})\sDet {\cal
U}=(\sDet {\cal A})\sDet \widetilde{{\cal A}}\;.
\eeq
Note that the
equality
\beq
\label{E45}
\sDet Q=\sDet\widetilde{Q}
\eeq
holds for any supermatrix $Q$,
\beq
\label{E46}
\widetilde{Q}_{i}^{\;j}=(-1)^{\varepsilon_i(\varepsilon_j+1)}Q^{j}{\;\!i}.
\eeq
Therefore  we prove the relation
\beq
\label{E47} J=(\sDet {\cal
A})^2 \eeq known as the factorization of the Grand Jacobian for
anticanonical transformations in the BV-formalism.

There exists a
possibility to express the factorization property of the Grand
Jacobian in terms of supermatrix ${\cal M}$ as well.
Indeed, let us use the presentation of
$J$ in the form \cite{Berezin}
\beq
\label{E48} J=\sDet H=\sDet ({\cal A}-{\cal
B}{\cal D}^{-1}{\cal C})\sDet {\cal D}^{-1}\;.
\eeq
Taking into
account that
\beq
\label{E49} {\cal A}={\cal M}-{\cal
K}\widetilde{{\cal M}}^{-1}{\cal N},\quad {\cal
D}^{-1}=\widetilde{{\cal M}},\quad {\cal B}={\cal K}\widetilde{{\cal
M}}^{-1}, \quad {\cal C}=\widetilde{{\cal M}}^{-1}{\cal N},
\eeq
we obtain
\beq
\label{E50} {\cal A}-{\cal B}{\cal D}^{-1}{\cal C}={\cal
M}, \eeq and \beq \label{E51} J=\sDet {\cal M}\sDet \widetilde{{\cal
M}}=(\sDet {\cal M})^2,
\eeq
the second presentation of the factorization property of the Grand Jacobian.

\section*{Acknowledgments}
\noindent
The work of Batalin and Tyutin  is supported  by the RFBR grant 20-02-00193.
The work of Lavrov is supported
by Ministry of Science and High Education of Russian Federation,
project FEWF-2020-0003.


\begin {thebibliography}{99}
\addtolength{\itemsep}{-8pt}

\bibitem{BV} I.A. Batalin, G.A. Vilkovisky, \textit{Gauge algebra and
quantization}, Phys. Lett. \textbf{B102} (1981) 27.

\bibitem{BV1} I.A. Batalin, G.A. Vilkovisky, \textit{Quantization of gauge
theories with linearly dependent generators}, Phys. Rev.
\textbf{D28} (1983)
2567.

\bibitem{brs1}
C. Becchi, A. Rouet, R. Stora, {\it The abelian Higgs Kibble Model,
unitarity of the $S$-operator}, Phys. Lett. {\bf B52} (1974)
344.

\bibitem{t}
I.V. Tyutin, {\it Gauge invariance in field theory and statistical
physics in operator formalism}, Lebedev Institute preprint  No.  39
(1975), arXiv:0812.0580 [hep-th].

\bibitem{Butt}
C. Buttin, {\it  Les de'rivations des champs de tenseurs et l'invariant
diffe'rentiel de Schouten}, C. R. Acad. Sci. Paris. Ser. A-B {\bf 269} (1969) 87.


\bibitem{VLT}
B.L. Voronov, P.M. Lavrov, I.V. Tyutin,
{\it Canonical transformations and gauge dependence
in general gauge theories},
Sov. J. Nucl. Phys. {\bf 36} (1982) 292.

\bibitem{LT}
P.M. Lavrov, I.V. Tyutin,
{\it Effective action in general gauge theories},
Yad. Fiz. {\bf 41} (1985) 1658.

\bibitem{BL}
I.A. Batalin, P.M. Lavrov,
{\it Closed description of arbitrariness in resolving quantum master equation},
Phys. Lett. {\bf B758} (2016) 54,

\bibitem{BLT}
I.A. Batalin, P.M. Lavrov, I.V.Tyutin,
{\it Finite anticanonical transformations in field-antifield formalism},
Eur. Phys. J. {\bf C75} (2015) 270.

\bibitem{BV2} I.A. Batalin, G.A. Vilkovisky,
\textit{Closure of the gauge algebra,generalized Lie algebra equations
and feynman rules},
Nucl. Phys. \textbf{B234} (1984) 106.

\bibitem{DeWitt}
B.S. DeWitt, \textit{Dynamical theory of groups and fields},
(Gordon and Breach, 1965).

\bibitem{Berezin}
F.A. Berezin, {\it Introduction to superanalysis},
Dordrecht-Boston, MA: D. Reidel, 1987.



\end{thebibliography}
\end{document}